\documentclass[a4paper]{IEEEtran}
\usepackage{cite}
\usepackage{amsmath,amssymb,amsfonts}
\usepackage{graphicx}
\usepackage{textcomp,nicefrac}
\usepackage{colortbl}
\usepackage{tikz}
\usepackage{subfig}
\usepackage[T1]{fontenc}
\usepackage{float}

\begin{document}

\title{Proton radiographs using position-sensitive silicon detectors and high-resolution scintillators}

\author{J.A. Briz, A.N. Nerio, C. Ballesteros, M.J.G. Borge, P. Martínez, A. Perea, V.G. Távora, O. Tengblad, M. Ciemala, A. Maj, P. Olko, W. Parol, A. Pedracka, B. Sowicki, M. Zieblinski, and E. Nácher.
\thanks{This work has been supported by the PRONTO-CM B2017/BMD-3888 project funded by Comunidad de Madrid (Spain). The research leading to these results has received funding from the European Union's Horizon 2020 research and innovation programme under grant agreement no. 654002 (ENSAR2) and grant agreement No [730983] (INSPIRE). This work has been partly supported by the Spanish Funding Agency for Research (AEI) through the PID2019-104390GB-I00 and PID2019-104714GB-C21 projects. A.N. Nerio acknowledges the fundings from the Erasmus Mundus Joint Master Degree on Nuclear Physics co-funded by the Erasmus+ Programme of the European Union.}
\thanks{J.A. Briz, A.N. Nerio, C. Ballesteros, M.J.G. Borge, P. Martínez, A. Perea, V.G. Távora and O. Tengblad are with Instituto de Estructura de la Materia (CSIC), 28006 Madrid, Spain (e-mail: jose.briz@csic.es).}
\thanks{A. Maj, P. Olko, M. Zieblinski, B. Sowicki, M. Ciemala, W. Parol and A. Pedracka are with Instytut Fizyki Jądrowej PAN, Krakow (Poland).}
\thanks{E. Nácher is with Instituto de Física Corpuscular (CSIC-Univ. Valencia), Valencia (Spain).}}

\maketitle

\begin{abstract}
Proton therapy is a cancer treatment technique currently in growth worldwide. It offers advantages with respect to conventional X-ray and $\gamma$-ray radiotherapy, in particular, a better control of the dose deposition allowing to reach a higher conformity in the treatments. Therefore, it causes less damage to the surrounding healthy tissue and less secondary effects. However, in order to take full advantage of its potential, improvements in treatment planning and dose verification are required. A new prototype of proton Computed Tomography scanner is proposed to design more accurate and precise treatment plans for proton therapy. Here, results obtained from an experiment performed using a 100-MeV proton beam at the CCB facility in Krakow (Poland) are presented. Proton radiographs of PMMA samples of 50-mm thickness with spatial patterns in aluminum were taken. Their properties were studied, including reproduction of the dimensions, spatial resolution and sensitivity to different materials. They demonstrate the capabilities of the system to produce images with protons. Structures of up to 2 mm are nicely resolved and the sensitivity of the system was enough to distinguish thicknesses of 10 mm of aluminum or PMMA. This constitutes a first step to validate the device as a proton radiography scanner previous to the future tests as a proton CT scanner.
\end{abstract}

\begin{IEEEkeywords}
proton therapy, proton radiograph, proton Computed Tomography, particle tracking, DSSD, silicon detectors, LaBr3, scintillation detectors.
\end{IEEEkeywords}

\section{Introduction}
\label{sec:introduction}
\begin{figure}[t]
\centerline{\includegraphics[width=\linewidth]{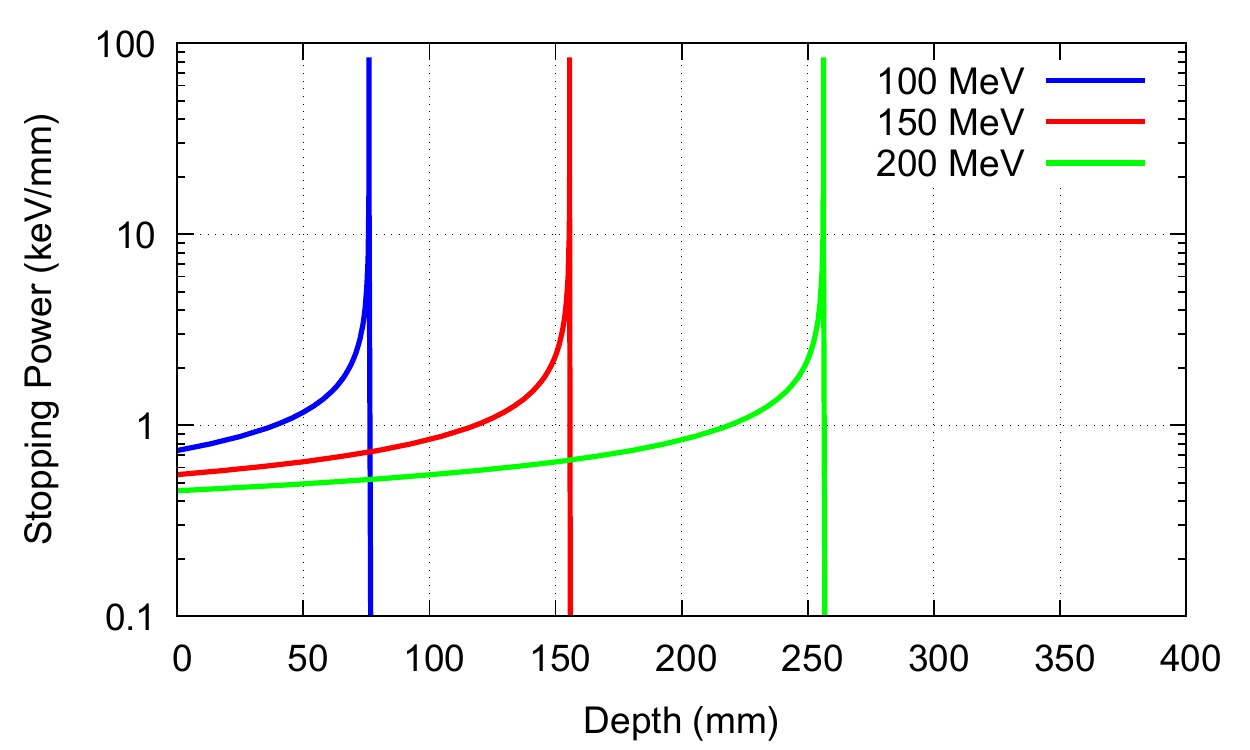}}
\caption{Bragg curves of protons with 100, 150, and 200 MeV in liquid water obtained from \cite{nucleonica}. It is shown as the energy is deposited mainly in the last few millimeters of the proton path, where the stopping power grows dramatically. Also shown is how one can control the location of the dose deposited by modifying the proton beam energy as the proton range varies with energy.}
\label{figBragg}
\end{figure}

\IEEEPARstart{P}{roton} therapy is a promising cancer treatment technique that is still far from being exploited to its full potential despite being already in use. Its main advantage with respect to conventional radiotherapy with X-rays or $\gamma$-rays is a more conformed dose deposition which can improve significantly the ratio of dose applied to the tumor with respect to the surrounding healthy tissue. Due to the properties of the slowing down process of protons in matter, they deposit most of their kinetic energy at the end of their path. This effect is shown in Fig.~\ref{figBragg} where the stopping power of protons of 100, 150, and 200 MeV in liquid water is represented as a function of the depth. These curves are usually referred to as Bragg curves. Protons deposit the majority of their kinetic energy at the end of their path where the stopping power grows remarkably. This feature is interesting for tumor treatment as it allows us to deposit energy at a given depth in the patient. The depth can be controlled by modifying the energy of the proton beam as shown in Fig.\ref{figBragg}. However, this property also implies that slight variations in the patient anatomy or inaccuracies estimating the proton range can be dramatic as a significant fraction of energy might be wrongly deposited in healthy tissue. Therefore, accurate and precise treatment plans and a system for in vivo dose verification during treatment are mandatory to extract the full potential of this technique.

Currently, proton therapy treatment plans are obtained using X-ray CT images. These provide the so-called maps of Hounsfield Units (HU), defined as:
\begin{equation}
    HU = 1000 \times \frac{\mu_{\textrm{tissue}} - \mu_{\textrm{water}}}{\mu_{\textrm{water}} - \mu_{\textrm{air}}}
\end{equation}
which accounts for the linear attenuation coefficient of the tissue ($\mu_{\textrm{tissue}}$). However, the proton therapy treatment plans are based on maps of a different physical quantity, the Relative Stopping Powers (RSP), implying that the HU maps have to be transformed into RSP maps. The conversion process is not straightforward and induces large uncertainties that, depending on the tissue, can cause uncertainties in proton ranges as large as 5 $\%$ in the abdomen and 11 $\%$ in the head, see Ref.~\cite{Joh18} and references therein.

Proton Computed Tomography (pCT) is an imaging modality that overcomes this difficulty by using the same particle for the planning and the treatment. In this way, the interaction with matter is the same in both cases and it is possible to obtain the RSP directly. This can reduce the uncertainties in proton ranges, ideally, below 1$\%$. Another advantage of the pCT technique is the reduced dose applied to the patient, being of the order of 1.3-1.4 mGy in contrast with the 30-50 mGy which are typical values needed to obtain an X-ray CT image \cite{Joh18}.

In this framework, research and development activities are on-going to construct a prototype of pCT scanner. As the first stage in this process, two experiments have been performed, one at the Centro de MicroAnálisis de Materiales (CMAM) in Madrid with proton beams of 10 MeV and, recently, an experiment with proton beams in the energy range of 90-120 MeV. The details of the former experiment can be seen in Ref.~\cite{procANIMMA}. In this article, the results on 2D radiographs from the last measurement are reported.

\section{Experiment}

The experiment was performed at Cyclotron Centre Bronowice (CCB) facility belonging to the Institute of Nuclear Physics PAN in Krakow (Poland). The IBA Proteus C235 cyclotron of the facility produced a 230-MeV proton beam that was degraded to the energy of interest (90-120 MeV) with an energy spread of 0.6$\%$. The beam spot size was of approximately 5-mm radius, and the beam intensity was between 500 pA - 1 nA depending on the beam energy.

\begin{figure}[ht]
\centerline{\includegraphics[width=\linewidth]{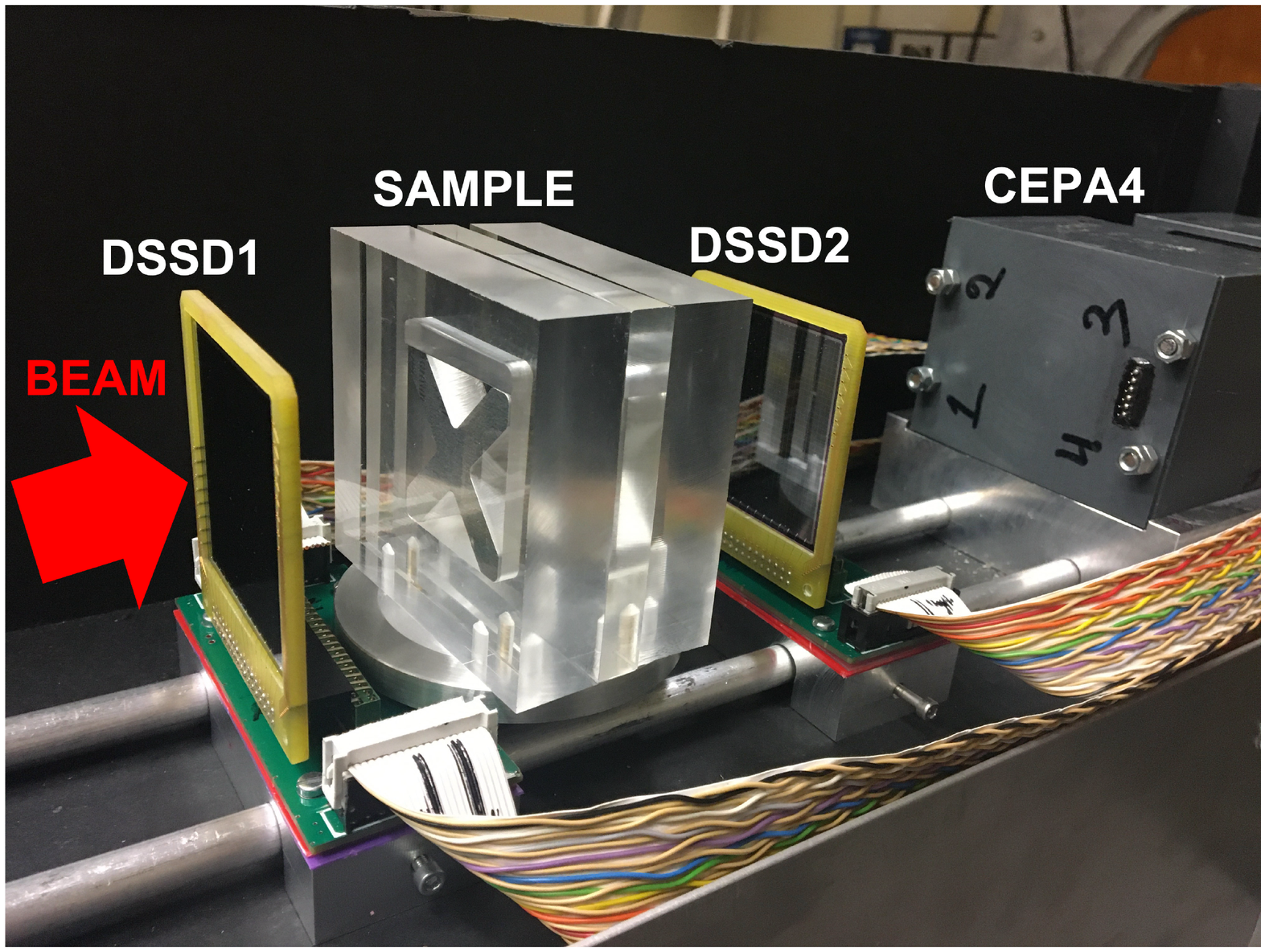}}
\caption{Experimental setup used to obtain the proton radiographs at the CCB facility. The proton CT scanner prototype is formed by two Double-sided Silicon Strip Detectors (DSSD) for proton tracking and the CEPA4 detector as residual-energy detector.}
\label{figSetup}
\end{figure}

The beam impinged on a 25-$\mu$m thick titanium target placed at the end of the beam line kept under vacuum. The Coulomb scattering occurring in the target reduced the initial beam intensity as a function of the scattering angle. The experimental setup shown in Fig.~\ref{figSetup} was centered at 12.5$^\circ$ with respect to the incident beam direction. In this way, the experimental setup was fully illuminated with the proton beam during the measurement. The experimental setup included two Double-sided Silicon Strip Detectors (DSSD) of 1-mm thickness, used as proton tracker. Each detector has a front surface of approximately $50\times50$ mm$^2$ with 16 vertical metal contacts on the front side and 16 horizontal on the back side to collect the charge produced by the charged particle traversing the active silicon layer. Like this, a mesh of $16\times16$ pixels of $3\times3$ mm$^2$ was defined and it provided the incident (at DSSD1) and outgoing (at DSSD2) positions of the proton at both planes. The two DSSDs were symmetrically placed at a distance of 70 mm to the central plane of the sample. A residual-energy detector is needed to fully stop the protons and, in this way, determine the energy they lost in the sample to be imaged. This detector was the CEPA4 detector \cite{CEPA4}, an array of 4 phoswich units, each of them formed by coupling a 4-cm-long LaBr$_3$(Ce) crystal and a 6-cm-long LaCl$_3$(Ce) crystal to a common photomultiplier. The front area of each phoswich unit is $27\times27$ mm$^2$. 

\begin{figure}[ht]
\includegraphics[width=\linewidth]{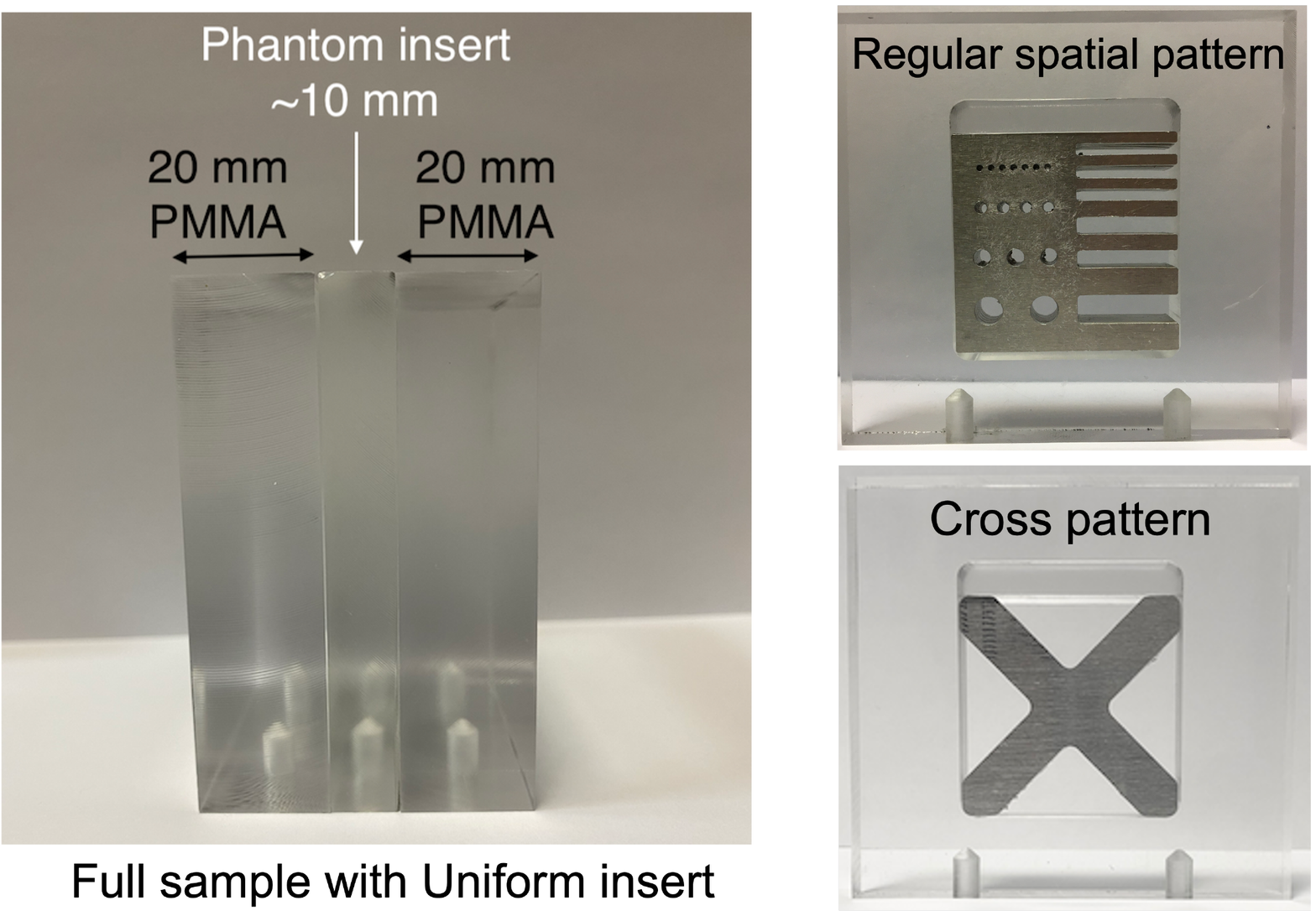}
\caption{Samples used in the experiment. Left image shows a lateral view of the general structure of the samples. They were formed by 3 pieces of Polymethyl methacrylate (PMMA) of a density $\rho_{\textrm{PMMA}}$=1.18 g/cm$^3$. The thicknesses of the first and last pieces of PMMA were 20 mm and they were always used. The internal piece labelled as ``phantom insert'' has a thickness of 10 mm and three options were possible: the uniform of PMMA (left), the regular spatial pattern (right-top), and the cross pattern (right-bottom). Both patterns were made of aluminum ($\rho_{\textrm{Al}}$=2.7 g/cm$^3$). The radiographs of the three samples were taken simply by exchanging the middle layer.}
\label{figSamples}
\end{figure}

Three different samples were used during the experiment, which are shown in Fig.~\ref{figSamples}. On the left side, the uniform sample formed by three pieces of Polymethyl methacrylate (PMMA, $\rho_{\textrm{PMMA}}$=1.18 g/cm$^3$) is shown. The first and last 20-mm PMMA pieces were used in all the measurements. Three inserts were placed as the middle layer and exchanged from measurement to measurement. The three available inserts were a uniform block of PMMA, a regular spatial pattern shown at the right-top side of Fig.~\ref{figSamples}, and a cross pattern insert shown at the right-bottom. The images of the three samples were obtained by exchanging the middle piece of 10 mm thickness labelled as ``Phantom insert'' in the left side image. 

\section{Data Analysis}

The energy calibrations of the detectors of the experimental setup were performed with proton beams of 95, 100, and 120 MeV energy and the sample with the uniform insert as shown in the left image of Fig.~\ref{figSamples}. In this way, three calibration points were obtained on each detector. Monte Carlo simulations of the experiment were performed using Geant4 \cite{Geant4} including a very detailed geometry of the setup. They provided us with the values of energy deposited in the different detectors for the three proton beam energies. 

A procedure of data filtering was followed to guarantee the proper identification of protons and the removal of electronic noise and background signals. First, at each DSSD detector, only events where a simultaneous signal was collected at both sides, one of a front strip (vertical) and one from a back (horizontal) strip with similar deposited energy were accepted as good events corresponding to a proton traversing the detector. In this way, the pixel hit in the DSSD was identified by the crossing of both triggered strips. A procedure to recover data from three strips of the DSSD2 detector, which were not collecting charge, was performed. It was based on the lower amplitude signal collected in the neighboring strips of those not working properly. Thanks to the different amount of charge collected, those events could be identified and assigned to the neighboring not-working strip. In these cases, the value of energy deposited was taken from the coincident strip of the other side (front or back, depending on the case) that was working in a proper way. Finally, the coincidence condition between the three detectors with certain energy restrictions were imposed to guarantee that the proton hit both the tracking and residual-energy detectors.

\subsection{Image reconstruction procedure}

The radiograph was obtained at the central plane of the sample. Due to the symmetrical distance of the two tracking detectors with respect to that central plane, the X,Y coordinates at the image plane were determined by simply averaging the X and Y coordinates at both detector planes: (x$_1$,y$_1$) in DSSD1 and (x$_2$,y$_2$) in DSSD2. In this way, the trajectory of the proton was assumed to be a straight line between the pixel hit in both detectors. The pixelation available at each detector plane is the one provided by the tracking detectors, which is a total of 256 pixels organized in a mesh of $16\times16$ strips of $3\times3$ mm$^2$ each. If no further data treatment was performed, the maximum granularity of the image formed would be $32\times32$ pixels resulting from the averaging of the coordinates of both detectors pixels. In order to generate images with higher granularity, the following procedure was applied. The impact position at each detector pixel was chosen randomly by generating a uniform random number in both X and Y axes. In this way, the counts assigned to each pixel were uniformly distributed along the full pixel area. The image has a surface of 48 x 48 mm$^2$, given by the frontal area of the DSSDs, which was divided in $128\times128$ pixels. For each traversing proton, the coordinates at the image plane were calculated as the average of those at each DSSD detector, (x$_{av}$,y$_{av}$)=((x$_1$+x$_2$)/2,(y$_1$+y$_2$)/2). At this point of the image, the total energy deposited in the three detectors was accumulated. The total accumulated energy at each pixel for all the detected protons was divided by the number of protons counted on that pixel to obtain the final image. In this way, an image of average energy deposited per proton was obtained. That image is called the proton radiograph from this device.

\section{Discussion}

The procedure previously described was used to obtain the radiographs of the three samples studied. The results obtained for the two samples with aluminum spatial patterns are presented in the following. The measurement with the uniform sample was only used for the calibration of the detectors.
\begin{figure}[ht]
\includegraphics[width=\linewidth]{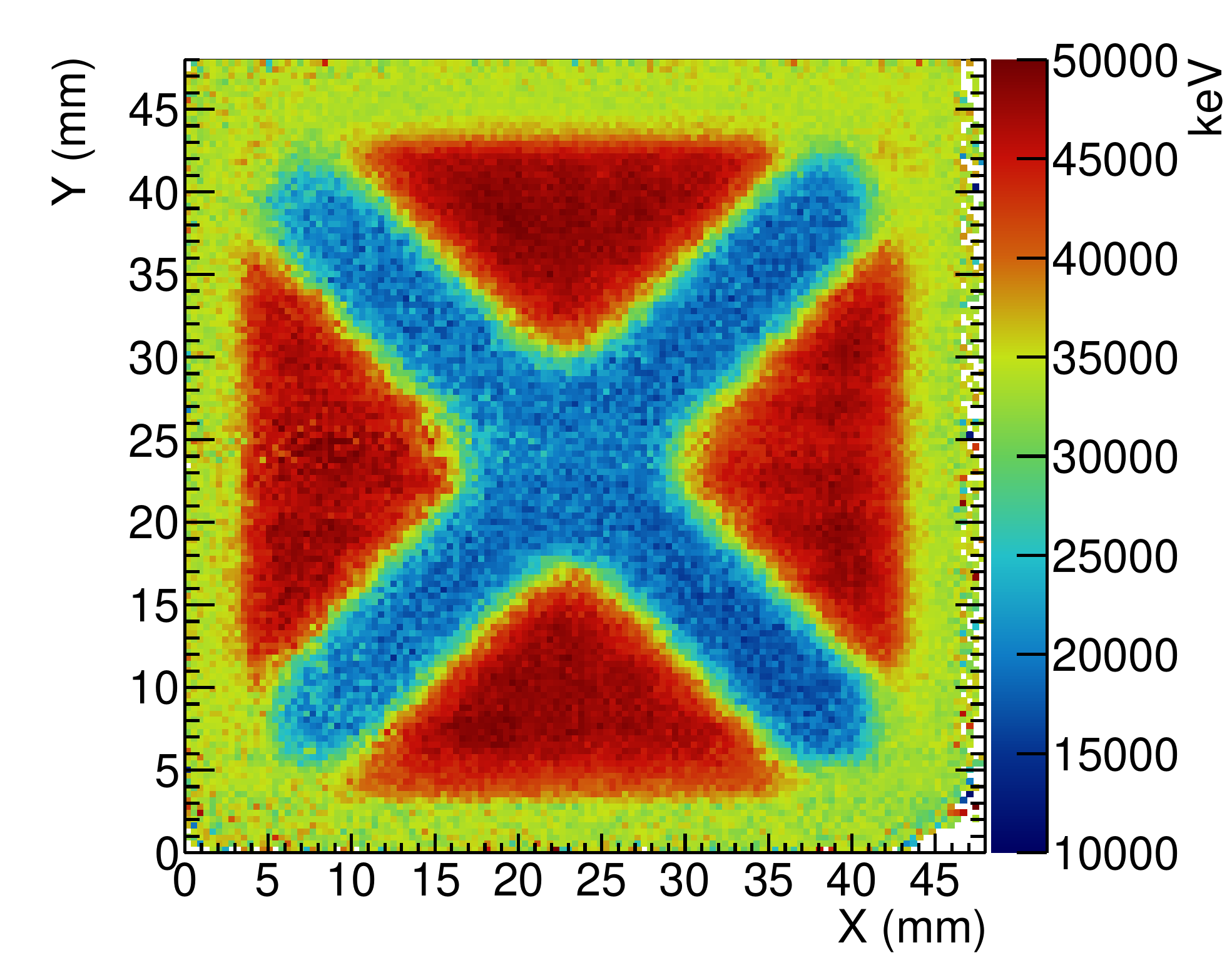}
\caption{Experimental image obtained with the cross-shaped pattern sample. The image is a two-dimensional map of energy detected in the scanner (the sum of energy deposited in DSSD1, DSSD2 and CEPA4 detectors). The position on the image plane was obtained using the hit positions on DSSD1 and DSSD2, and it was expressed in the X,Y plane in units of mm.}
\label{figCrossImage}
\end{figure}

\subsection{Cross pattern sample}

The image shown in Fig.~\ref{figCrossImage} corresponding to the cross pattern shown in Fig.~\ref{figSamples} was obtained using the procedure described previously. Three main regions are clearly distinguished, the one in red corresponds to the zone where the insert has no material, which is called the ``Air'' region; the greenish zone is referred to as ``PMMA'' region and it is where PMMA material is located as the square frame around the pattern; and the blue zone which is the cross pattern made of aluminum, that is called the ``Aluminum'' region. One has to keep in mind that the cross pattern piece is sandwiched between two PMMA pieces of 20 mm thickness each. So, for example, protons passing by the "Air" region are effectively traversing 40 mm of PMMA plus 10 mm of air with a much lower density ($\rho_{\textrm{air}}$=1.2$\times$10$^{-3}$ g/cm$^3$).

\begin{figure}[ht]
\centerline{\includegraphics[width=\linewidth]{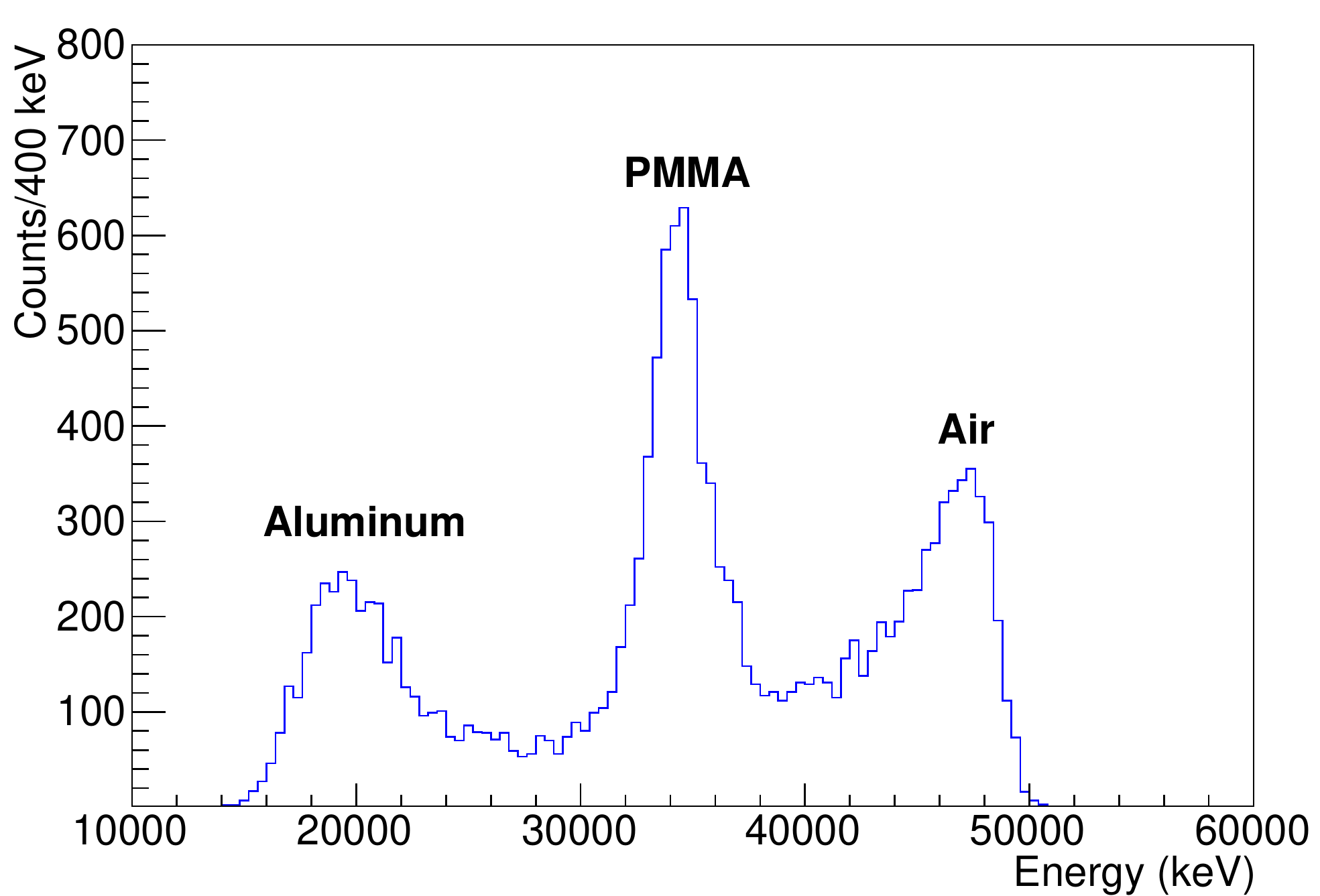}}
\caption{Histogram of the energy values obtained from the image pixels of the cross pattern sample shown in Fig.~\ref{figCrossImage}. The counts on the vertical axis make reference to the number of pixels in the image with an energy value in the energy range covered by the 400-keV-wide bin. Three main peaks are observed corresponding to the three main regions visible in the image. They correspond to the Aluminum, PMMA, and Air regions of the sample insert shown in Fig.~\ref{figSamples}.}
\label{figCrossHist}
\end{figure}

In order to better understand the properties of the image, an energy histogram was generated. The value of accumulated energy stored at each pixel was used to fill the energy histogram. The histogram is filled with one count per pixel, that is $128\times128=16384$ counts in the full spectrum. The result is shown in Fig.~\ref{figCrossHist}. The spectrum was obtained to estimate the power of the device to distinguish between different materials. The histogram shows the three resolved structures corresponding to regions of different materials: Aluminum with $\rho_{\text{Al}}=2.7$ g/cm$^3$, PMMA with $\rho_{\text{PMMA}}=1.18$ g/cm$^3$ and Air with $\rho_{\text{air}}=1.2\times 10^{-3}$ g/cm$^3$ of 10 mm thickness approximately each, can be resolved. A fit to Gaussian functions was done for each of the three main peaks observed, resulting in energy values of 19.6(18) MeV for the peak corresponding to Aluminum, 34.3(15) MeV for that of PMMA, and 47.1(12) MeV for the Air region. It is worth recalling here that those regions correspond to the sample insert that contained the pattern and that was placed between two PMMA blocks of 20-mm thickness. So this means that from the histogram it can be said that 100-MeV protons traversing 40-mm of PMMA, corresponding to the ``Air'' region, are losing about 53 MeV (100-47.1 MeV) and those traversing 50-mm of PMMA, corresponding to the ``PMMA'' region, are losing almost 66 MeV (100-34.3 MeV).

\begin{figure}[t]
\includegraphics[width=\linewidth]{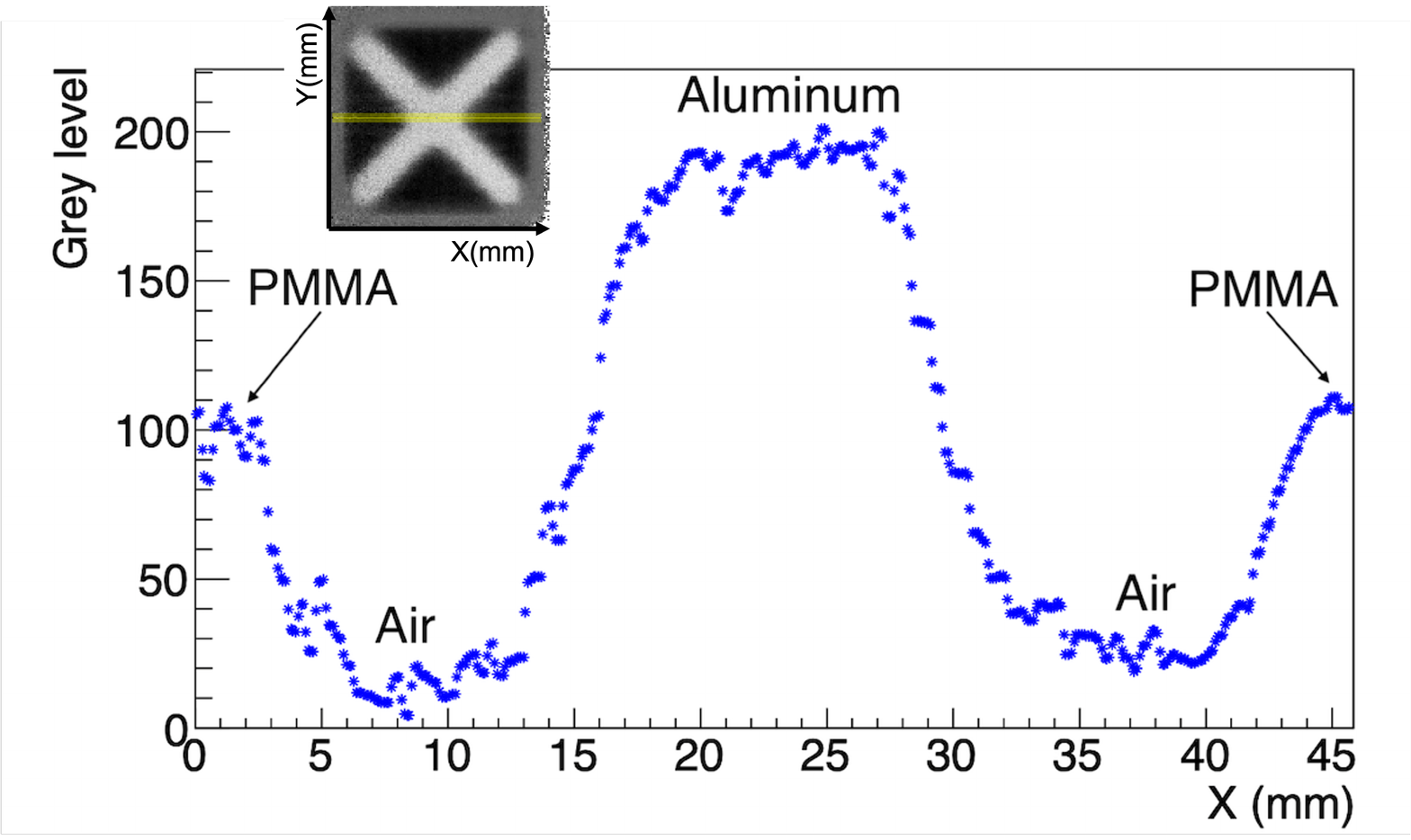}
\caption{Grey level profile of the image obtained for the cross pattern sample in the region marked with yellow in the inset image at the top-left corner of the plot. Regions of different materials are distinguished and labelled in the plot.}
\label{figCrossGLP}
\end{figure}

\begin{table*}[ht]
    \centering
    \caption{Measurements in the cross pattern sample.}
    \label{tab:Dimensions_Cross}
    \setlength{\tabcolsep}{10pt}
        \begin{tabular}{c >{\columncolor[gray]{0.8}}c c | c c | c c}
             \hline
               & Real value & Fit order & FWHM & $\left|\frac{\text{FWHM} - \text{Real}}{\text{Real}}\right|$ & FWTM & $\left|\frac{\text{FWTM} - \text{Real}}{\text{Real}}\right|$ \\
             Dimension & (mm) & $a_5$ & (mm) & (\%) & (mm) & (\%) \\
             \hline
             \hline
            W	 & 39.70(5) & 12	 & 34.93(14)	 & 12	 & 38.60(15)	 &  3 \\
            H	 & 39.85(5) & 12	 & 35.15(15)	 & 12	 & 38.85(16)	 &  3 \\
            c1	 & 14.45(5) &  8	 & 14.07(10)	 &  3	 & 16.34(11)	 & 13 \\
            d1	 & 54.30(5) & 10	 & 48.7(2)  	 & 10	 & 54.9 (2) 	 &  1 \\
            w1	 &  9.00(5) &  6	 &  7.31(5) 	 & 19	 &  8.93(6) 	 &  1 \\
            h1	 & 12.60(5) &  4	 & 14.04(7) 	 & 17	 & 18.95(10)	 &  1 \\
            b1	 & 26.15(5) &  8	 & 24.87(8) 	 &  5	 & 28.90(9) 	 & 11 \\
         \end{tabular}
        \newline
        \newline
        The column in grey corresponds to the dimensions of the sample as measured directly from the piece. Dimensions are described in Fig. \ref{figCross_Dim}.
\end{table*}
The obtained images were converted to greyscale to be analysed using the ImageJ software \cite{imagej}. One-dimensional profiles were extracted from the image by selecting a region of interest (ROI). If the region contains several rows or columns of pixels, the profile shows the average values in the selected region. Fig.~\ref{figCrossGLP} shows the profile of the region indicated in yellow on the inset at the top-left of the image obtained for the cross pattern sample.

\begin{figure}[t]
\centerline{
\includegraphics[width=0.5\linewidth]{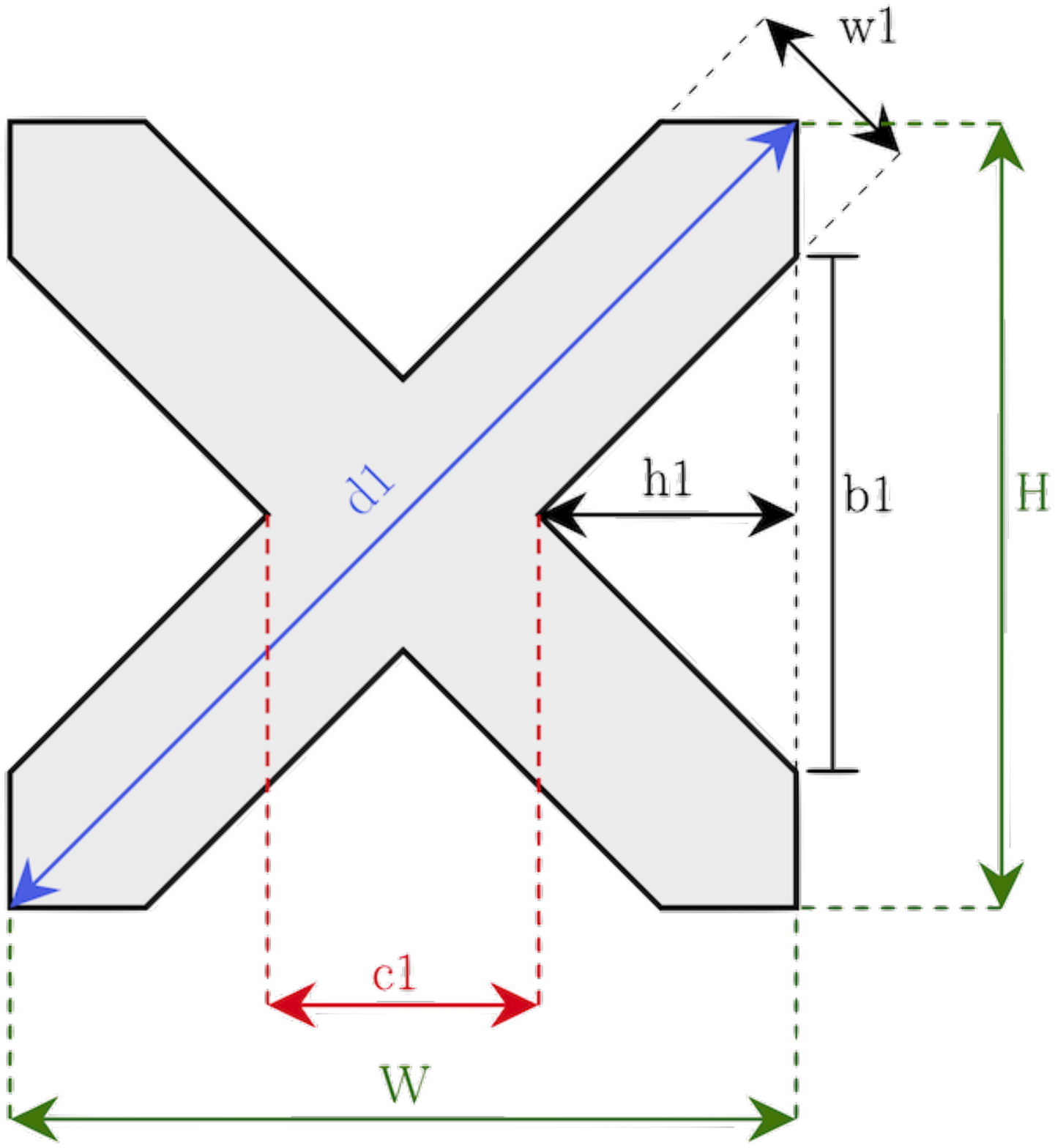}}
\caption{Main dimensions studied of the image of the cross pattern. The values for those dimensions were obtained by fitting the grey level profiles of the image at the region of interest to super-Gaussian functions, see Eq.~\ref{eq:supgaus}. The results are given in Table~\ref{tab:Dimensions_Cross}.}
\label{figCross_Dim}
\end{figure}

The dimensions were estimated by fitting the profiles to generalized Gaussian functions to obtain the best possible fit to the shape of the line profile. The general expression for a generalized Gaussian function, also known as super-Gaussian, is given in Eq.~\ref{eq:supgaus}, where a linear background has been added. The parameters a$_0$ and a$_1$ describe the linear background, a$_2$ is the standard deviation, $\sigma$, of the function, a$_ 3$ describes the total area of the function above the linear background considered, a$_4$ is the centroid (central position) and the $a_5$ parameter is the super-Gaussian order. The last parameter was conveniently modified for each region to be fitted. In general, larger order indicates a more extended flat-top region in the shape of the function.
\begin{equation}
f(x)=a_0+a_1 x + \frac{a_3}{\sqrt{2\pi}a_2}\cdot exp\left(-\frac{1}{2}\left(\frac{x-a_4}{a_2}\right)^{a_5}\right)
\label{eq:supgaus}
\end{equation}
From the fitted function, the Full Width at Half-Maximum (FWHM) and at Tenth-Maximum (FWTM) were determined. The dimensions indicated in Fig.~\ref{figCross_Dim} were obtained and the resulting values are listed in Table~\ref{tab:Dimensions_Cross} compared with the values obtained from measurements on the actual piece. The FWTM values generally showed lower deviations from the measured values than the FWHM giving deviations lower than 3$\%$ for most cases.

\subsection{Regular spatial pattern sample}

\begin{figure}[ht]
\includegraphics[width=\linewidth]{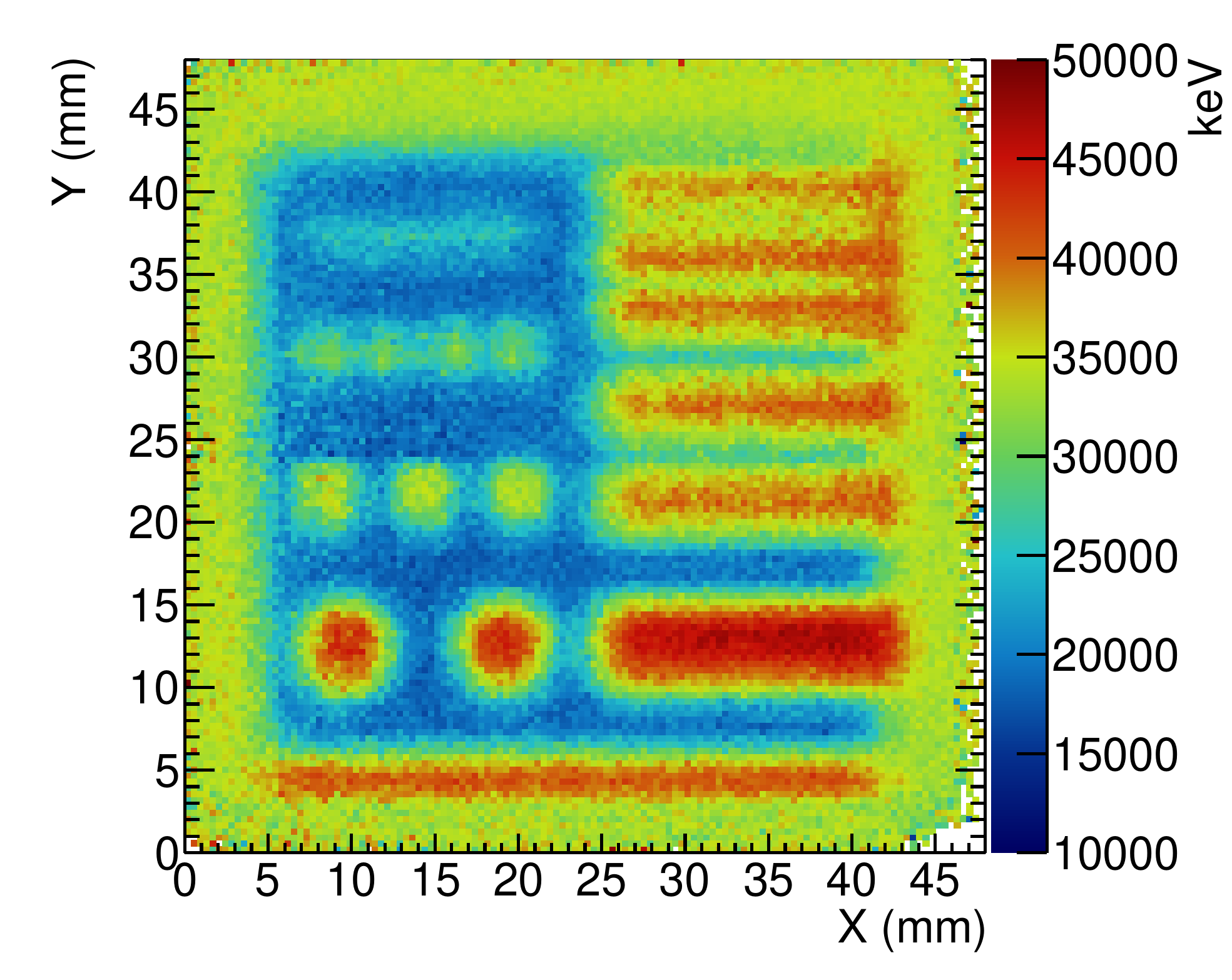}
\caption{
Experimental image obtained with the regular spatial pattern sample. The image is a two-dimensional map of energy detected in the scanner (sum of energy deposited in DSSD1, DSSD2, and CEPA4 detectors). The position on the image plane was obtained using the hit positions on DSSD1 and DSSD2, and it was expressed in the X,Y plane in units of mm.
}
\label{figDerenzoImage}
\end{figure}

\begin{figure}[ht]
\centerline{\includegraphics[width=0.63\linewidth]{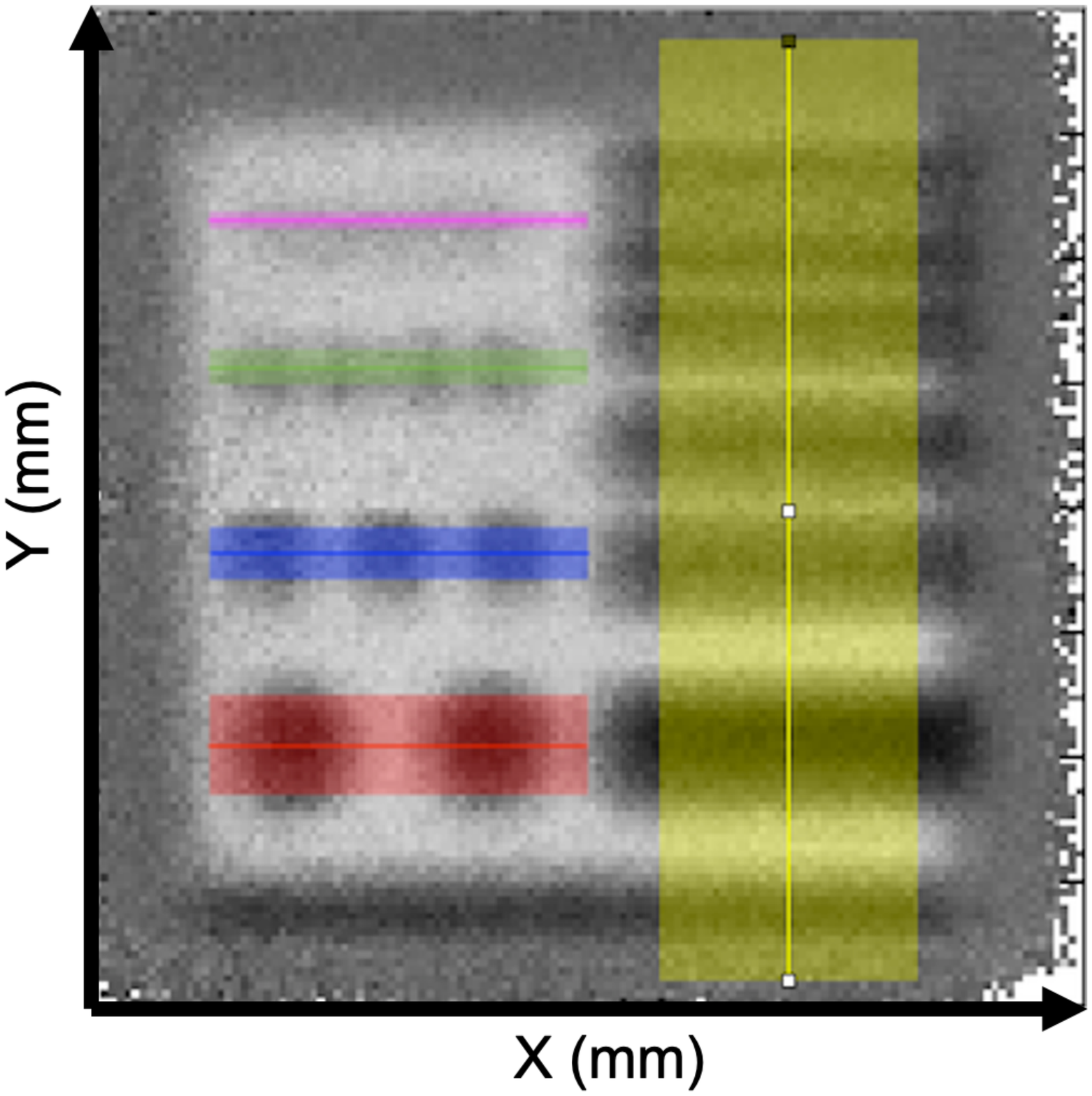}}
\caption{Regions Of Interest (ROIs) defined on the greyscale image obtained for the regular pattern sample shown at the right-hand side of Fig.~ \ref{figDerenzoImage} but here represented in greyscale. The profiles corresponding to the ROIs here indicated are given in Figs.~\ref{figSlitsDerenzo} and \ref{figDerenzoProfiles}.}
\label{figROIsDerenzo}
\end{figure}

A sample containing a regular spatial pattern of holes and slits made on a 10-mm-thick aluminum piece, which is shown in Fig.~\ref{figSamples}, was studied to explore the spatial resolution achievable in proton radiographs with this device. On the left side, it contains four rows of circular holes with spacing equal to its diameter. Two circles of 5-mm diameter form the bottom row; three circles of 3-mm diameter form the second one; the third row contains four circles of 2-mm diameter; and seven circles of 1-mm diameter form the top one. On the right side of the sample, a regular pattern of slits of different thicknesses is built. From bottom to top, it has two aluminum slits of 5 mm separated by 5 mm, two slits of 3 mm spaced by 3-mm gaps and then three slits of 2 mm separated by spacings of 2 mm. Fig.~\ref{figDerenzoImage} shows the image obtained for a 1-hour measurement with this sample with a proton beam of 100 MeV. The region of aluminum material is clearly seen in blue tones, while the holes of larger dimensions are also clearly distinguished in red tones. The smaller the size of the pattern is, the lower the contrast between the aluminum and air regions it is. This can be seen when comparing 5-mm holes with those of 3 or 2 mm where the internal part of the holes is shown in yellowish or greenish tones instead of red as expected.

\begin{figure}[ht]
\centerline{\includegraphics[width=\linewidth]{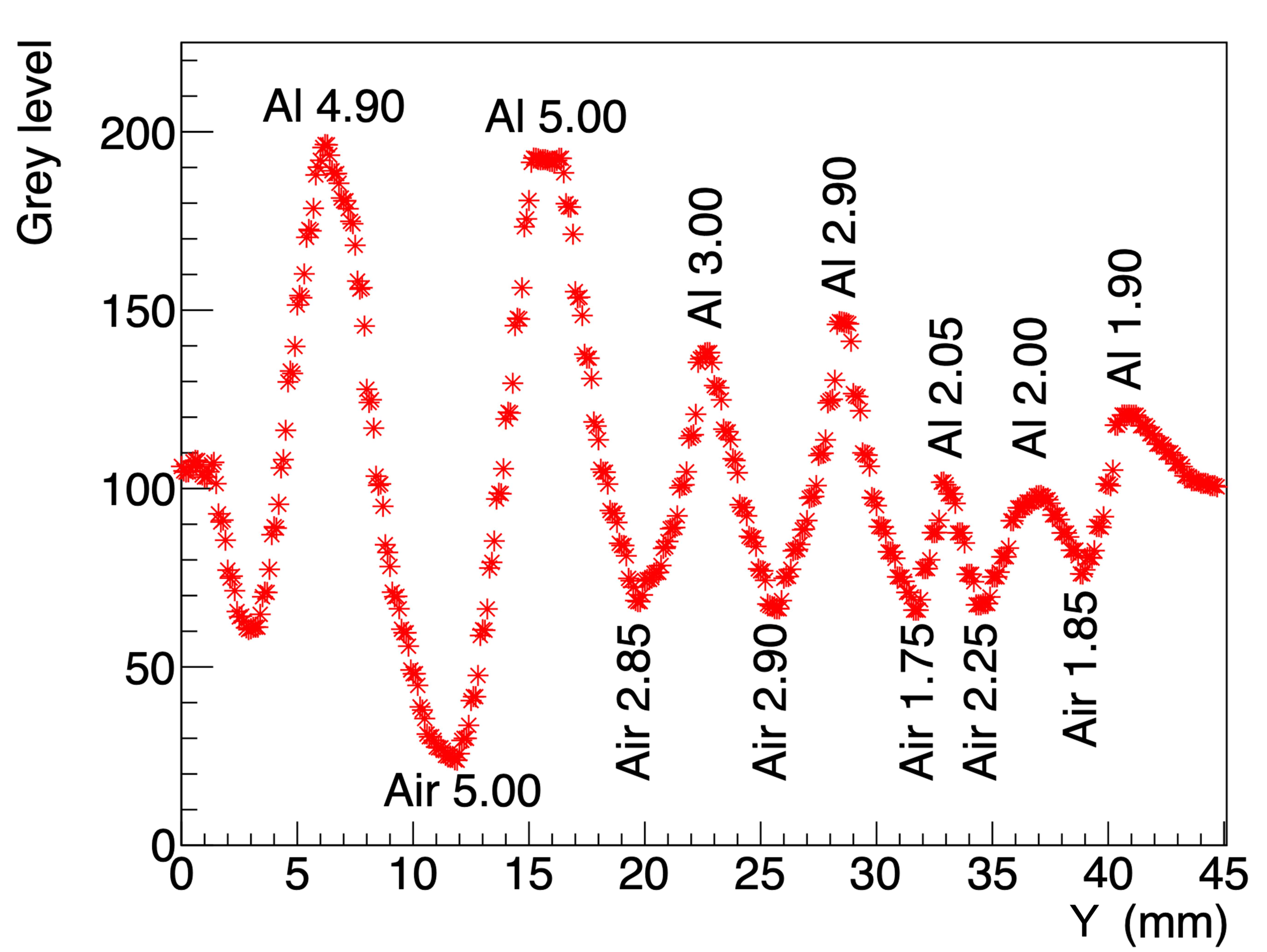}}
\caption{Profile obtained at the ROI marked in yellow in Fig.~\ref{figROIsDerenzo}. Labels indicate the material and size in mm of each region. Values given are those obtained from a direct measurement of the actual piece. Values of the dimensions of the slits mentioned in the text are rounded for simplicity in the discussion, so they have to be considered approximations.}
\label{figSlitsDerenzo}
\end{figure}

\begin{figure}[ht]
\centerline{\includegraphics[width=\linewidth]{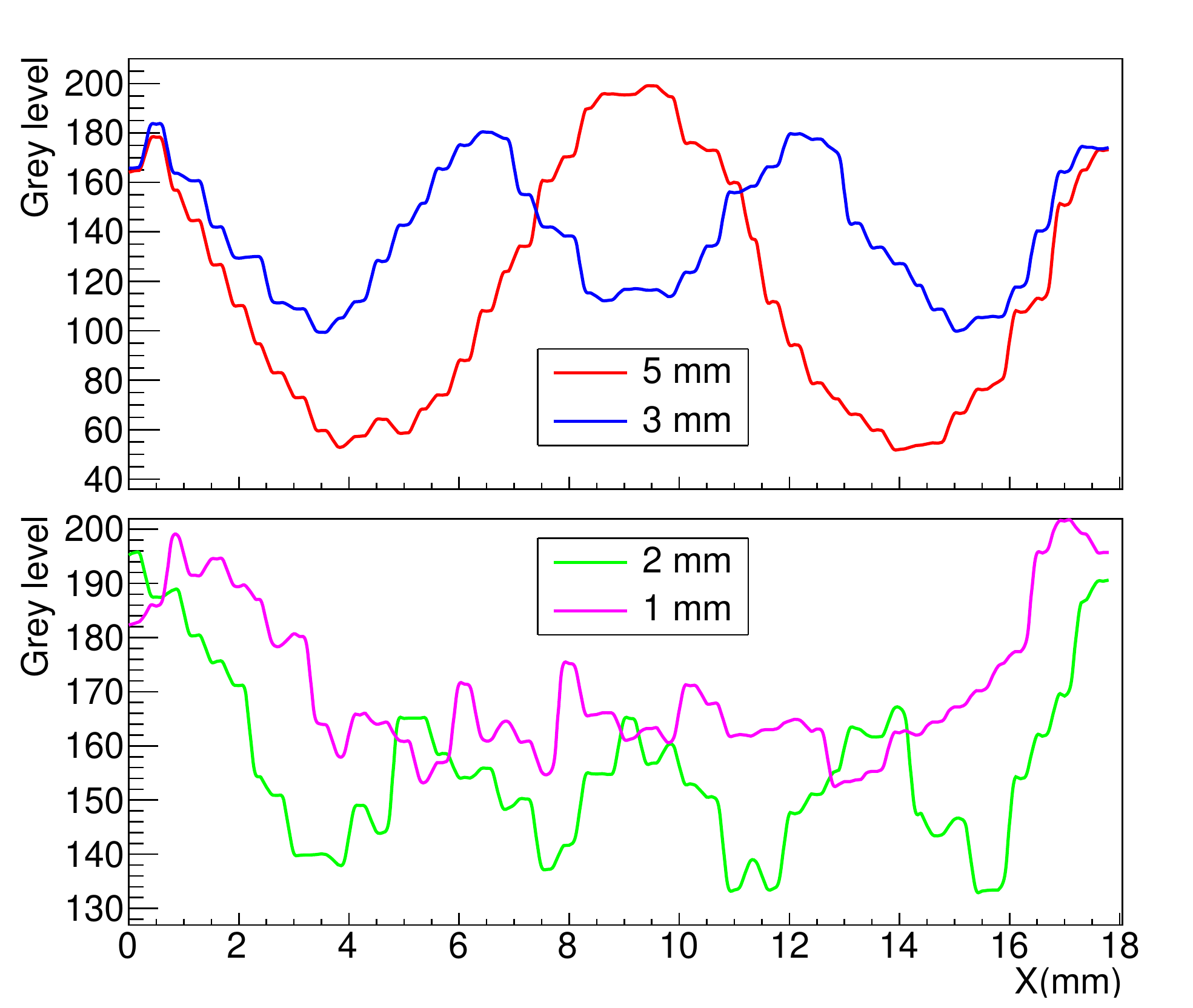}}
\caption{Comparison of one-dimensional profiles of the regular spatial pattern sample at the four rows of holes of different sizes. Top pad shows the curves for the 5- and 3-mm rows and bottom pad for those of 2- and 1-mm. The color of the lines indicates the regions of interest marked in Fig.~\ref{figROIsDerenzo}. Two minima corresponding to the two holes of 5 mm, three minima for the row of 3 holes of 3 mm, and four for the row of 4 holes of 2 mm diameter are resolved, while the ones of 1 mm can not be clearly distinguished.}
\label{figDerenzoProfiles}
\end{figure}

In order to see more details of the sample, five regions of interest (ROIs) were selected which are shown in Fig.~\ref{figROIsDerenzo}. The ROI indicated in yellow corresponds to the zone where the slits are located
and its profile is shown in Fig.~\ref{figSlitsDerenzo}. The profile shows different maxima, which correspond to the aluminum slits, and several minima corresponding to the empty zones between the aluminum slits of the sample insert. The two first slits of $\approx$5 mm are clearly visible and seen with high contrast. Those of $\approx$3 mm are seen next with lower contrast but well resolved, and, finally, the three slits of 2 mm are also resolved with even lower contrast. This profile shows us that the spatial resolution of the system is better than 2 mm.

The profiles obtained for the rest of the ROIs indicated with different colors in Fig.~\ref{figROIsDerenzo} are shown in Fig.~\ref{figDerenzoProfiles}. The structures of 5-, 3- and 2-mm diameters and spacings are resolved. The two minima in the row of two holes of 5-mm diameter are clearly observed in red. Similarly, the three minima for the row of three holes of 3-mm diameter shown in blue are well resolved. For the row of 2-mm diameter shown in green, the contrast is lower, but the structure of four holes can still be observed and resolved. On the contrary, the profile for 1-mm holes shown in magenta, does not show a clear pattern with seven minima as expected. In conclusion, the spatial resolution of the radiographs is better than 2 mm but not reaching 1 mm.

\section{Conclusions}
The performance of a proton Computed Tomography scanner is under study. As a first stage, tests of the device to perform proton radiographs have been performed. An experiment carried out at Cyclotron Centre Bronowice facility in Krakow (Poland) using proton beams in the energy range from 90 to 120 MeV is presented here. A proton radiography scanner formed by two DSSD detectors for proton tracking and a scintillator array for residual-energy determination was used. Energy calibrations of the detectors were performed using proton beams of 95, 100, and 120 MeV and detailed Monte Carlo simulations of the experimental setup. Different samples made of PMMA and spatial patterns made from aluminum inserted in the PMMA matrix were studied. The images acquired for a cross pattern sample and a regular spatial pattern sample have been studied. The dimensions of the cross pattern were determined from the image using fits to super-Gaussian functions and, in general, they reproduce the dimensions of the actual piece under study. In addition, the energy histogram showed three peaks located at well differentiated energies which indicates that the device has sensitivity to distinguish a thickness of 10-mm of these materials (aluminum, PMMA and air). In continuation of this work, a detailed study in order to determine the limits of the sensitivity to different RSPs will be performed. From the image of the sample with regular spatial pattern, the holes of 5, 3, and 2 mm were resolved and only those of 1 mm were not properly identified. Therefore, the spatial resolution of the device is estimated to be better than 2 mm. 

The results presented here constitute the first step to validate the system as a proton radiography scanner. The next steps will be to improve the algorithm of image reconstruction, specially on how the counts at each pixel are distributed within the pixel area, and to perform the data analysis of tomography scans performed by taking several proton radiographs at different rotation angles of the sample. The images will be used to perform the tomographic reconstruction of the object. Further tests of the device as a proton Computed Tomography scanner will be performed. 

\section*{Acknowledgments}
This work has been supported by the PRONTO-CM B2017/BMD-3888 project funded by Comunidad de Madrid (Spain). The research leading to these results has received funding from the European Union's Horizon 2020 research and innovation programme under grant agreement no. 654002 (ENSAR2) and grant agreement No [730983] (INSPIRE). This work has been partly supported by the Spanish Funding Agency for Research (AEI) through the PID2019-104390GB-I00 and PID2019-104714GB-C21 projects.
A.N. Nerio acknowledges the fundings from the Erasmus Mundus Joint Master Degree on Nuclear Physics co-funded by the Erasmus+ Programme of the European Union.

\end{document}